\documentclass[twoside]{dis07}
\usepackage[latin1]{inputenc}
\usepackage[dvips]{graphicx,epsfig,color}
\usepackage{wrapfig,rotating}
\usepackage{amssymb,amsmath,array}

\begin{document}

\newcommand{\beq}[1]{
\begin{equation}\label{#1}}
\newcommand{\eeq}{\end{equation}}
\newcommand{\bea}[1]{
\begin{eqnarray}\label{#1}}
\newcommand{\eea}{\end{eqnarray}}

\title{Vector Meson production from NLL BFKL}

\author{Dmitry Yu. Ivanov$^1$ and Alessandro Papa$^2$
%
%
\vspace{.3cm}\\
%
1- Sobolev Institute of Mathematics \\ 630090 Novosibirsk - Russia
%
\vspace{.1cm}\\
2- Dipartimento di Fisica, Universit\`a della Calabria \\
        and INFN - Gruppo Collegato di Cosenza \\
        I-87036 Rende - Italy\\ }

\maketitle

\begin{abstract}
The amplitude for the forward electroproduction of two light
vector mesons can be written completely within perturbative QCD in
the Regge limit with next-to-leading accuracy, thus providing the
first example of a physical application of the BFKL approach at
the next-to-leading order. We study in the case of equal photon
virtualities the main systematic effects, by considering a
different representation of the amplitude and different
optimization methods of the perturbative series.
\end{abstract}

\section{Introduction}

In the BFKL approach~\cite{BFKL}, both in the leading logarithmic
approximation (LLA), which means resummation of all terms
$(\alpha_s\ln(s))^n$, and in the next-to-leading approximation
(NLA), which means resummation of all terms
$\alpha_s(\alpha_s\ln(s))^n$, the amplitude for a large-$s$ hard
collision process can be written as the convolution of the Green's
function of two interacting Reggeized gluons with the impact
factors of the colliding particles.

The Green's function is determined through the BFKL equation. The
kernel of the BFKL equation is known now both in the
forward~\cite{NLA-kernel} and in the non-forward~\cite{FF05}
cases. On the other side, impact factors are known with NLA
accuracy in a few cases: colliding partons~\cite{partonIF},
forward jet production~\cite{BCV03} and forward transition from a
virtual photon $\gamma^*$ to a light neutral vector meson
$V=\rho^0, \omega, \phi$~\cite{IKP04}. The most important impact
factor for phenomenology, the $\gamma^* \to \gamma^*$ impact
factor, is calling for a rather long calculation, which seems to
be close to completion now~\cite{gammaIF,Cha}.

The $\gamma^* \to V$ forward impact factor can be used together
with the NLA BFKL forward Green's function to build, completely
within perturbative QCD and with NLA accuracy, the amplitude of
the $\gamma^* \gamma^* \to V V$ reaction. This amplitude provides
us with an ideal theoretical laboratory for the investigation of
several open questions in the BFKL approach. Besides, this process
can be studied experimentally at the future at ILC, see Refs.~\cite{EPSW1}.


\section{Representations of the NLA amplitude}

The process under consideration
is the
production of two light vector mesons ($V=\rho^0, \omega, \phi$)
in the collision of two virtual photons, $ \gamma^*(p) \:
\gamma^*(p')\to V(p_1) \:V(p_2)$. Here, neglecting the meson mass
$m_V$, $p_1$ and $p_2$ are taken as Sudakov vectors satisfying
$p_1^2=p_2^2=0$ and $2(p_1 p_2)=s$; the virtual photon momenta are
instead $p =\alpha p_1-Q_1^2/(\alpha s) p_2$ and $p'=\alpha^\prime
p_2-Q_2^2/(\alpha^\prime s) p_1$, so that the photon virtualities
turn to be $p^2=-Q_1^2$ and $(p')^2=-Q_2^2$. We consider the
kinematics when $ s\gg Q^2_{1,2}\gg \Lambda^2_{QCD}$ and
$\alpha=1+Q_2^2/s+{\cal O}(s^{-2})$, $\alpha^\prime
=1+Q_1^2/s+{\cal O}(s^{-2})$. In this case vector mesons are
produced by longitudinally polarized photons in the longitudinally
polarized state~\cite{IKP04}. Other helicity amplitudes are power
suppressed, with a suppression factor $\sim m_V/Q_{1,2}$. We will
discuss here the amplitude of the forward scattering, i.e. when
the transverse momenta of produced $V$ mesons are zero or when the
variable $t=(p_1-p)^2$ takes its maximal value
$t_0=-Q_1^2Q_2^2/s+{\cal O}(s^{-2})$.

The NLA forward amplitude can be written as a spectral
decomposition on the basis of eigenfunctions of the LLA BFKL
kernel:

\[
\frac{\mbox{Im}_s\left( {\cal A}_{\mathrm exp}\right)}{D_1D_2}
=\frac{s}{(2\pi)^2} \int\limits^{+\infty}_{-\infty} d\nu
\left(\frac{s}{s_0}\right)^{\bar \alpha_s(\mu_R) \chi(\nu)+\bar
\alpha_s^2(\mu_R) \left( \bar
\chi(\nu)+\frac{\beta_0}{8N_c}\chi(\nu)\left[-\chi(\nu)+\frac{10}{3}
\right] \right)} \alpha_s^2(\mu_R) c_1(\nu)c_2(\nu)
\]
\beq{amplnlaE} \times\! \left[1+\bar \alpha_s(\mu_R)
\left(\frac{c^{(1)}_1(\nu)}{c_1(\nu)}
+\frac{c^{(1)}_2(\nu)}{c_2(\nu)}\right) +\bar
\alpha_s^2(\mu_R)\ln\left(\frac{s}{s_0}\right)
\frac{\beta_0}{8N_c}\chi(\nu)\left(
i\frac{d\ln(\frac{c_1(\nu)}{c_2(\nu)})}{d\nu}+2\ln(\mu_R^2)
\right)\right]. \eeq

Here the bulk of NLA kernel corrections are exponentiated, ${\bar
\alpha_s}=\alpha_s N_c/\pi$ and $D_{1,2}=-4\pi e_q f_V/(N_c
Q_{1,2})$, where $f_V$ is the meson dimensional coupling constant
($f_{\rho}\approx 200\, \rm{ MeV}$) and $e_q$ should be replaced
by $e/\sqrt{2}$, $e/(3\sqrt{2})$ and $-e/3$ for the case of
$\rho^0$, $\omega$ and $\phi$ meson production, respectively. Two
scales enter the expression~(\ref{amplnlaE}), the renormalization
scale $\mu_R$ and the scale for energy $s_0$.

Alternatively, the amplitude can be expressed as a series:
\bea{series} \frac{Q_1Q_2}{D_1 D_2}\frac{\mbox{Im}_s ({\cal
A}_{\mathrm series})}{s} &=& \frac{1}{(2\pi)^2} \alpha_s(\mu_R)^2
\label{honest_NLA} \\ & \times & \biggl[ b_0
+\sum_{n=1}^{\infty}\bar \alpha_s(\mu_R)^n \, b_n \,
\biggl(\ln\left(\frac{s}{s_0}\right)^n   +
d_n(s_0,\mu_R)\ln\left(\frac{s}{s_0}\right)^{n-1}     \biggr)
\biggr]\;. \nonumber \eea The $b_n$ coefficients are determined by
the kernel and the impact factors in LLA, while the $d_n$
coefficients depend also on the NLA corrections to the kernel and
to the impact factors. We refer to Ref.~\cite{IP06} for the
details of the derivation and for the definition of the functions
entering these expressions.

\section{Numerical results}

\begin{figure}
\includegraphics[width=0.49\textwidth]{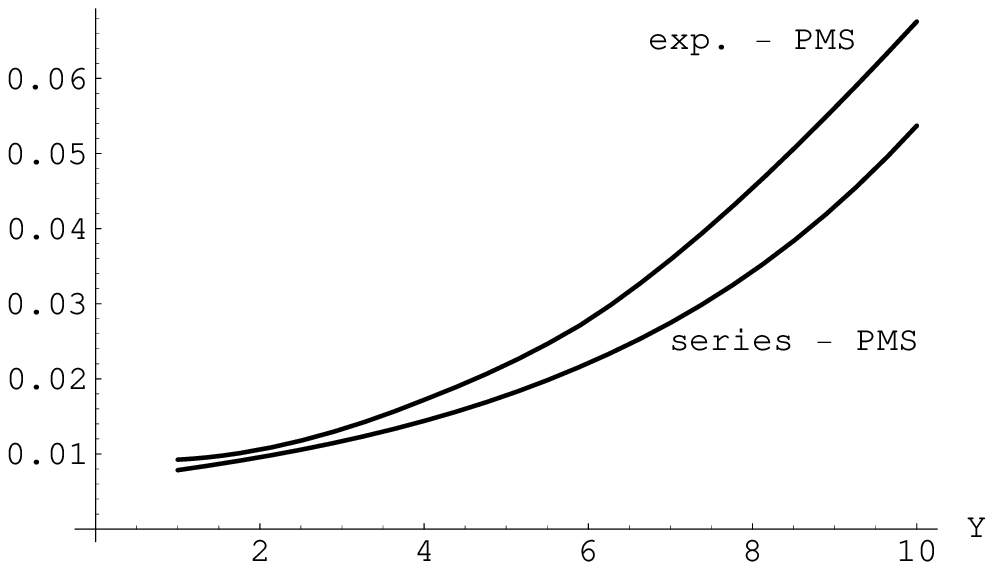}
\includegraphics[width=0.49\textwidth]{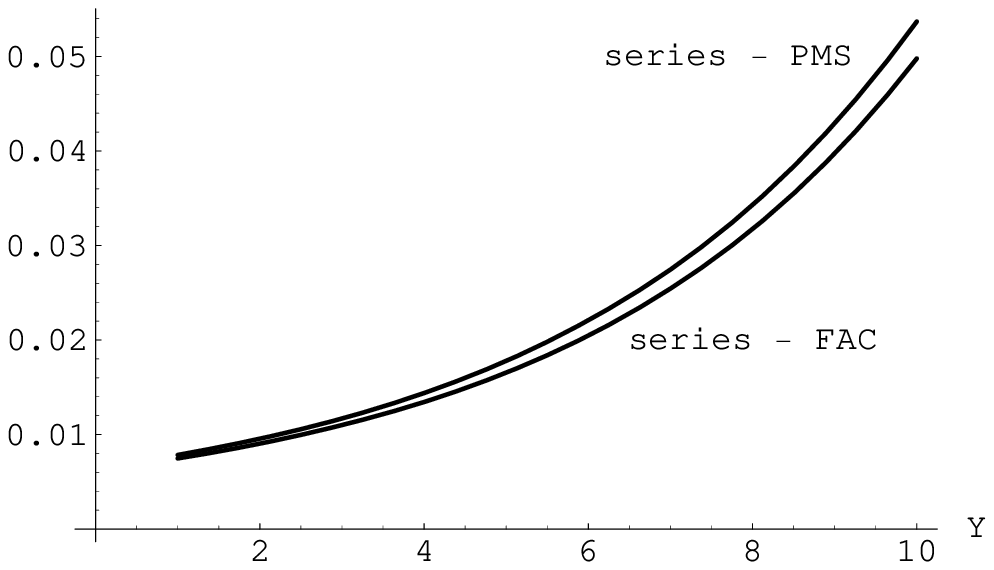}
\caption[]{\small $\mbox{Im}_s ({\cal A})Q^2/(s \, D_1 D_2)$ as a
function of $Y$ at $Q^2$=24 GeV$^2$ ($n_f=5$): (left) series
representation with PMS and ``exponentiated'' representation with
PMS, (right) series representation with PMS and with FAC.}
\label{comp1}
\end{figure}

\begin{figure}
\includegraphics[width=0.49\textwidth]{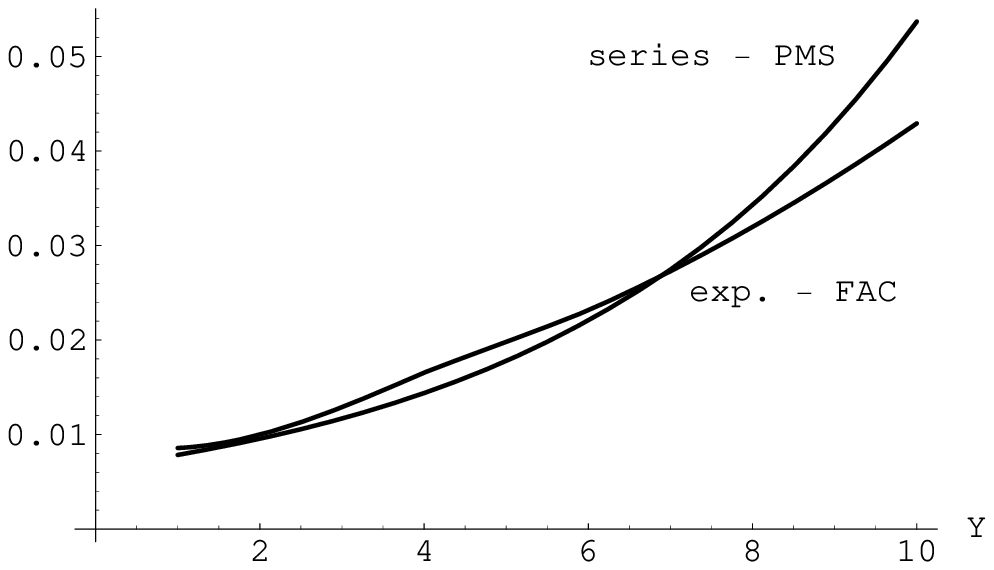}
\includegraphics[width=0.49\textwidth]{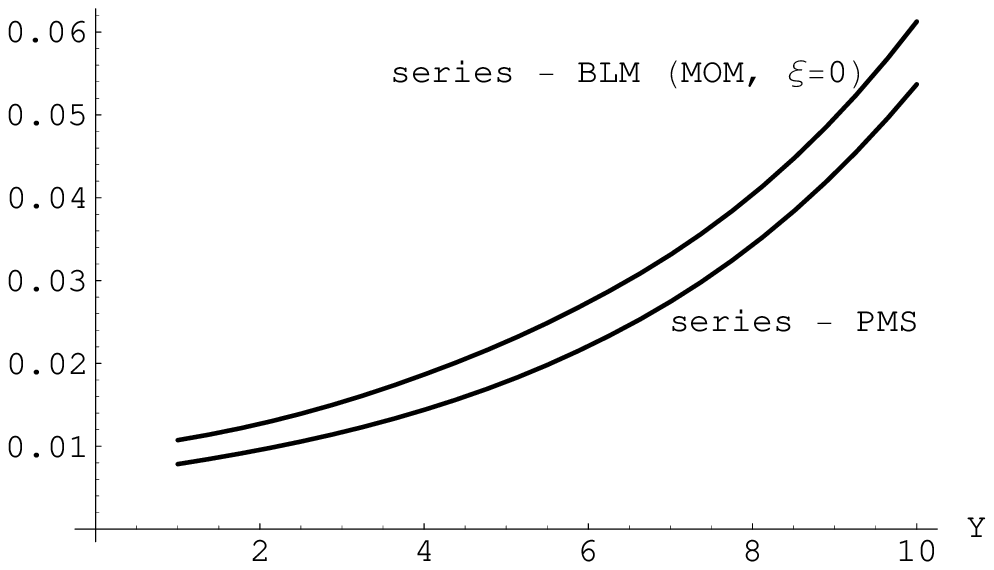}
\caption[]{\small $\mbox{Im}_s ({\cal A})Q^2/(s \, D_1 D_2)$ as a
function of $Y$ at $Q^2$=24 GeV$^2$ ($n_f=5$): (left) series
representation with PMS and ``exponentiated'' representation with
FAC, (right) series representation with PMS and with BLM.}
\label{comp2}
\end{figure}

In Ref.~\cite{IP06} we presented some numerical results for the
amplitude given in Eq.~(\ref{series}) for the $Q_1=Q_2\equiv Q$
kinematics, i.e. in the ``pure'' BFKL regime. We found that the
$d_n$ coefficients are negative and increasingly large in absolute
values as the perturbative order increases, making evident the
need of an optimization of the perturbative series. We adopted the
principle of minimal sensitivity (PMS)~\cite{Stevenson}, by
requiring the minimal sensitivity of the predictions to the change
of both the renormalization and the energy scales, $\mu_R$ and
$s_0$. We considered the amplitude for $Q^2$=24 GeV$^2$ and
$n_f=5$ and studied its sensitivity to variation of the parameters
$\mu_R$ and $Y_0=\ln(s_0/Q^2)$. We could see that for each value
of $Y=\ln(s/Q^2)$ there are quite large regions in $\mu_R$ and
$Y_0$ where the amplitude is practically independent on $\mu_R$
and $Y_0$ and we got for the amplitude a smooth behaviour in $Y$
(see the curve labeled ``series - PMS'' in Figs.~\ref{comp1}
and~\ref{comp2}).
The optimal values turned out to be $\mu_R\simeq 10 Q$ and
$Y_0\simeq 2$, quite far from the kinematical values $\mu_R=Q$ and
$Y_0=0$. These ``unnatural'' values probably mimic large unknown
NNLA corrections.

As an estimation of the systematic effects in our determination,
we considered also the ``exponentiated'' representation of the
amplitude, Eq.~(\ref{amplnlaE}), and different optimization
methods. For more details on the following, see Ref.~\cite{IP}.

At first, we compare the series and the ``exponentiated''
determinations using in both case the PMS method. The optimal
values of $\mu_R$ and $Y_0$ for the ``exponentiated'' amplitude
are quite similar to those obtained in the case of the series
representation, with only a slight decrease of the optimal
$\mu_R$. Fig.~\ref{comp1} (left) shows that the two determinations
are in good agreement at the lower energies, but deviate
increasingly for large values of $Y$. It should be stressed,
however, that the applicability domain of the BFKL approach is
determined by the condition $\bar \alpha_s(\mu_R) Y \sim 1$ and,
for $Q^2$=24 GeV$^2$ and for the typical optimal values of
$\mu_R$, one gets from this condition $Y\sim 5$. Around this value
the discrepancy between the two determinations is within a few
percent.

As a second check, we changed the optimization method and applied
it both to the series and to the ``exponentiated'' representation.
The method considered is the fast apparent convergence (FAC)
method~\cite{Grun}, whose strategy, when applied to a usual
perturbative expansion, is to fix the renormalization scale to the
value for which the highest order correction term is exactly zero.
In our case, the application of the FAC method requires an
adaptation, for two reasons: the first is that we have two energy
parameters in the game, $\mu_R$ and $Y_0$, the second is that, if
only strict NLA corrections are taken, the amplitude does not
depend at all on these parameters. For details about the
application of this method, we refer to~\cite{IP}. Here, we merely
show the results: the FAC method applied to the series
representation (see Fig.~\ref{comp1} (right)) and to the
exponentiated representation (see Fig.~\ref{comp2} (left)) gives
results in nice agreement with those from the PMS method applied
to the series representation, over the whole energy range
considered.

Another popular optimization method is the
Brodsky-Lepage-Mackenzie (BLM) one~\cite{BLM}, which amounts to
perform a finite renormalization to a physical scheme and then to
choose the renormalization scale in order to remove the
$\beta_0$-dependent part. We applied this method only to the
series representation, Eq.~(\ref{series}). The result is compared
with the PMS method in Fig.~\ref{comp2} (right) (for details, see
Ref.~\cite{IP}).

\section*{Acknowledgments}

The work of D.I. was partially supported by grants
RFBR-05-02-16211, NSh-5362.2006.2.


\begin{footnotesize}


\end{footnotesize}


\end{document}